\def\be{\begin{equation}}
\def\ee{\end{equation}}

\documentclass[prl,preprint,superscriptaddress,showpacs]{revtex4}
\usepackage[dvips]{graphicx}
\usepackage{amssymb}

\begin{document}

\title{Relation between Rotational and Translational\\ 
Dynamic Heterogeneities in Water}
\date{\today}

\author{Marco G. Mazza}
\affiliation{Center for Polymer Studies and Department of Physics,
  Boston University, Boston, Massachusetts 02215, USA}
\author{Nicolas Giovambattista}
\affiliation{Department of Chemical Engineering, Princeton
 University,\\Princeton, New Jersey 08544-5263, USA}
\affiliation{Center for Polymer Studies and Department of Physics,
  Boston University, Boston, Massachusetts 02215, USA}
\author{Francis  W. Starr}
\affiliation{Department of Physics, Wesleyan University, Middletown,
  Connecticut 06459, USA} 
\author{ H. Eugene Stanley} 
\affiliation{Center for Polymer Studies and Department of Physics,
  Boston University, Boston, Massachusetts 02215, USA}

\begin{abstract}
We use molecular dynamics simulations to probe the rotational dynamics
of the SPC/E model of water for a range of temperatures down to $200$~K,
$13$~K above to the mode coupling temperature. We find that rotational
dynamics is spatially heterogeneous, i.e., there are clusters of
molecules that rotate significantly more than the average for a given
time interval, and we study the size and the temporal behavior of these
clusters. We find that the position of a rotational heterogeneity is
strongly correlated with the position of a translational heterogeneity,
and that the fraction of molecules belonging to both kinds of
heterogeneities increases with decreasing temperature.  We further find
that although the two types of heterogeneities are not identical, they
are related to the same physical picture.

\end{abstract}

\maketitle

Experiments and computer simulations have shown that dynamics in
supercooled liquids is spatially heterogeneous, i.e., one can identify
transient regions with relaxation times different by orders of magnitude
\cite{silledi}. Simulations have shown that the most mobile particles
tend to form clusters \cite{glotzer,harrowell}. Different theoretical
approaches have been developed to understand spatially heterogeneous
dynamics~\cite{AG,chandler}. In particular Adam and Gibbs (AG)
\cite{AG,debene} postulate the existence of cooperatively rearranging
regions (CRR) whose molecules change configuration independently of the
rest of the system. Molecular dynamics (MD)
simulations~\cite{AntonioNature,nico} have verified many of the
predictions of the AG theory, which has also been tested in other
systems~\cite{gebreVG}.

While there have been numerous studies of the heterogeneous nature of
the translational degrees of freedom (TDOF) in water, there are few
studies of the heterogeneous nature of the rotational degrees of freedom
(RDOF). Here we systematically study the rotational dynamics of water
and compare with the translational dynamics. Previous work for other
systems suggests that translationally mobile molecules may have enhanced
rotational mobility~\cite{keys,ribeiro}, and the characteristic times
for the RDOF are smaller than for the TDOF~\cite{matsui}. To this end we
perform MD simulations of a system of $N=1728$ water molecules
interacting with the extended simple point charge potential (SPC/E)
\cite{spce} for a range of temperatures from $350$~K down to $200$ at
the fixed density of $1~\text{g/cm}^3$; for each temperature we run two
independent trajectories to improve the statistics. Simulation details
are given in \cite{francislong}. 

We quantify the rotation of a molecule using the normalized polarization
vector $\hat{p}_i(t)$. In a time interval $[t,t+\delta t]$,
$\hat{p}_i(t)$ spans an angle
$\delta\theta\equiv\cos^{-1}\left(\,\hat{p}_i(t)
\cdot\hat{p}_i(t+\delta t)\right)$. We define a vector 
$\delta\vec{\varphi}_i(t)$ such that
$|\delta\vec{\varphi}_i(t)|=\delta\theta$ and its direction is given by
$\hat{p}_i(t)\times\hat{p}_i(t+\delta t)\,$. Thus the vector
$\vec{\varphi}_i(t)\equiv\int_0^t\delta\vec{\varphi}_i(t')dt'$ allows us
to define a trajectory in a $\varphi$-space representing the rotational
motion of molecule $i$. One can then associate a rotational mean square
displacement (RMSD) given by
\be
<\varphi^2(t)>\equiv\frac{1}{N}\sum_i|\vec{\varphi}_i(t)-\vec{\varphi}_i(0)|^2
\label{rmsd}
\ee
and a rotational diffusion coefficient 
\be
D_r\equiv\lim_{t\to\infty}\frac{1}{4tN}
\sum_{i=1}^N\langle\,|\vec{\varphi}_i(t)-
\vec{\varphi}_i(0)|^2\rangle\,. \label{dr}
\ee
The vector $\vec{\varphi}_i(t)$ is not bounded
to the unit sphere, since otherwise Eq.~(\ref{dr}) would give $D_r=0$.
Equations~(\ref{rmsd}) and (\ref{dr}) were applied in \cite{KKS} 
to study the rotational motion of a linear molecular
system. 

\begin{figure}
\begin{center}
\includegraphics[scale=0.65]{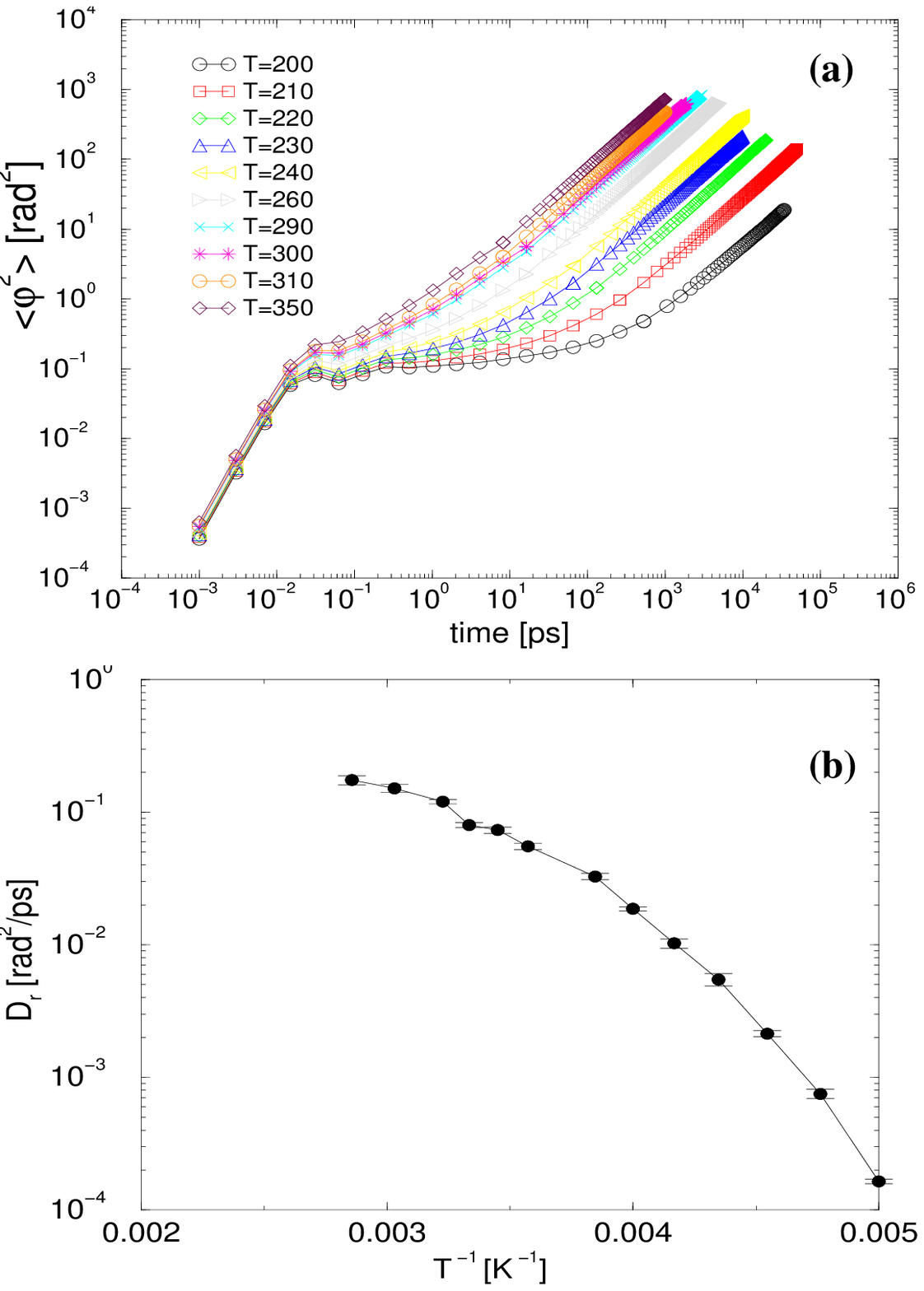}
\end{center}
\caption{(a) RMSD for a range of temperatures from $200$ to $350$ K. (b)
Rotational diffusivity as a function of $T^{-1}$.\label{r-msd}}
\end{figure}

Figures~\ref{r-msd}(a) and \ref{r-msd}(b) show the RMSD and the
temperature dependence of $D_r$, respectively. Similar to
Ref.~\cite{KKS} we observe: (i) the RMSD shows three different
regimes: a \emph{ballistic} regime, where $<\varphi^2>\propto t^2$, a
plateau or \emph{cage} regime, where molecules find themselves trapped in
the cage formed by their neighboring molecules, and finally a
\emph{diffusive} regime where $<\varphi^2>\propto t$, these three regimes
are analogous to those observed in studies of translational dynamics of
supercooled liquids. (ii) $D_r$ increases with $T$ with a non-Arrhenius
behavior. Note the oscillations in Fig.~\ref{r-msd}(a) present for times
of the order of $10^{-2}$~ps. These oscillations, not present for linear
molecules \cite{KKS}, correspond to the \emph{libration} (hindered
rotation) regime and occur at the same time as the oscillations observed
in the rotational correlation function of water in
\cite{rotscior}.

\begin{figure}
\begin{center}
\includegraphics[scale=0.6]{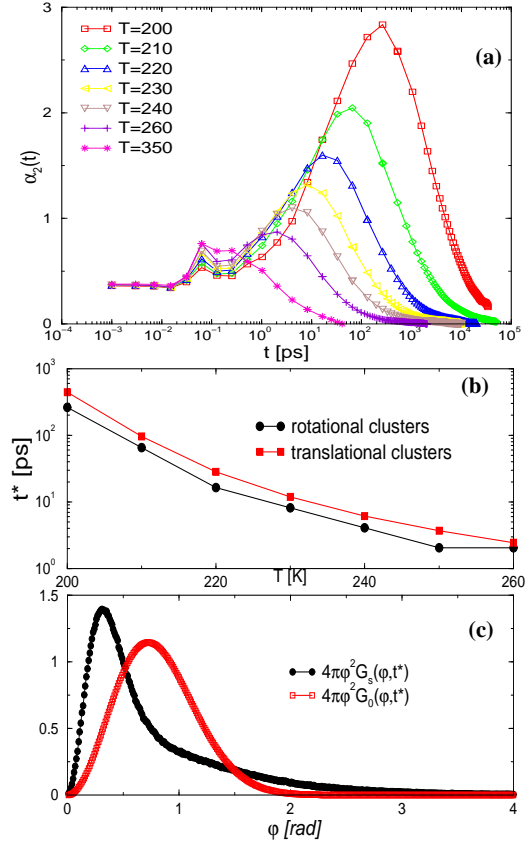}
\end{center}
\caption{(a) Non-Gaussian parameter $\alpha_2(t)$ for 
the range of temperatures indicated. (b) $T$-dependence of $t^*(T)$, the
time at which the maximum of $\alpha_2$ occurs. For comparison, we also
show the the value of $t^*$ obtained from the study of the TDOF in
\cite{nico}. (c) Self part of the van Hove
distribution function, $G_s(\varphi,t)$, for $\mathrm{T}=200$ K and
$t^*(200~\mathrm{K})\approx 1.05~\mathrm{ps}$, compared with the
Gaussian approximation, $G_0(\varphi,t)$, obtained using
$\langle\varphi^2(t^*)\rangle$ also at $\mathrm{T}=200$ K.\label{hove}}
\end{figure}

We introduce the rotational counterpart of the self part of the
time-dependent van Hove distribution function, $G_s(\varphi,t)$
\cite{McDonald},
\be
G_s(\varphi,t)\equiv\frac{1}{N}\sum_{i=1}^N\langle
\delta(|\vec{\varphi}_i(t)-\vec{\varphi}_i(0)|-\varphi)\rangle\,,
\ee
where $\langle\cdots\rangle$ represents average over
configurations. With this formalism we recover the usual interpretation
for $4\pi\varphi^2G_s(\varphi,t)$ as the probability of having a
molecule at time t with angular displacement $\varphi$. In other words,
that in the abstract $\varphi$-space, a molecule has moved to a distance
$\varphi$ from its position at $t=0$. For long times the diffusion
equation for $\vec{\varphi}_i(t)$ holds, and $G_s(\varphi,t)$ is a
Gaussian distribution
\be
G_0(\varphi,t)=\left[\frac{3}{2\pi\langle\varphi^2(t)\rangle}
\right]^{3/2}\exp\left[-3\varphi^2/2\langle\varphi^2(t)\rangle\right]\,.
\ee
The deviations of $G_s(\varphi,t)$ from $G_0(\varphi,t)$ can be
quantified by the non-Gaussian parameter \cite{alpha}
\be
\alpha_2(t)\equiv\frac{3}{5}\frac{\langle\varphi^4(t)\rangle}
{\langle\varphi^2(t)\rangle^2}-1\,.
\ee
Figure~\ref{hove}(a) shows $\alpha_2(t)$ for different temperatures.
$\alpha_2(t)$ shows a clear maximum at $t=t^*(T)$ which corresponds to
the beginning of the diffusive regime for the RDOF
[Fig.~\ref{r-msd}(a)]. We note that there is a small maximum at
$t\approx 10^{-2}$~ps: this is a consequence of the librational motion
as shown in Fig.~\ref{r-msd}(a) \cite{note1}. Figure~\ref{hove}(c) shows
$G_s(\varphi,t)$ and $G_0(\varphi,t)$ for $T=200$~K and $t=t^*(200
K)\approx 1.05$~ns. As in \cite{nico} we find that $G_s(\varphi,t^*)$
and $G_0(\varphi,t^*)$ intersect for large $\varphi$ at $\varphi^*$, and
$G_s(\varphi,t^*)$ shows a large tail where the fitted Gaussian
underestimate the angular motion of the molecules. Molecules with
$\varphi>\varphi^*$ can be considered with an angular displacement
higher than expected; this fraction $f\equiv\int_{\varphi^*}^\infty
4\pi\varphi^2G_s(\varphi,t)$ is found to be $\approx 13\%$, showing a
weak $T$-dependence.

In analogy to \cite{nico,donatiGlotzer,science3D}, we define the
rotational mobility of a molecule at a given time $t_0$ as the maximum
angular displacement in the interval $[t_0,t_0+\Delta t]$ of the oxygen
atom
\be
\Psi_i(t,\Delta t)\equiv\max\{|\vec{\varphi}_i(t+t_0)-\vec{\varphi}_i(t_0)|
\,,
\,t_0\le t\le t_0+\Delta t\}\,.
\ee
We focus our attention on the most rotationally mobile molecules and
explore the possibility that there exist clusters also among this
category of molecules. To facilitate comparison with the study
\cite{nico} of translational heterogeneities (TH), we select the $7\%$
of the most rotationally mobile molecules \cite{note2} and define a
cluster at time $t_0$ over an observation time $\Delta t$ as those
molecules whose nearest neighbor oxygen-oxygen (O-O) distance at time
$t_0$ is less than $0.315$ nm (first minimum of O-O radial distribution
function). We find that the rotational dynamics is spatially
heterogeneous, since these molecules form clusters, which we will call
rotational heterogeneities (RH).

\begin{figure}
\begin{center}
\includegraphics[scale=0.42]{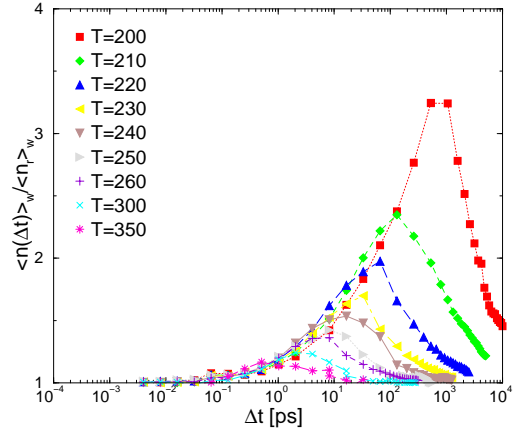}
\end{center}
\caption{Weight average cluster size for $\mathrm{T}=200~\text{K}-350$ K for
molecules belonging to the $7\%$ most rotationally mobile molecules. The
values are normalized to the random cluster contribution.
\label{clusize}}
\end{figure}

\begin{figure}
\begin{center}
\includegraphics[scale=0.9]{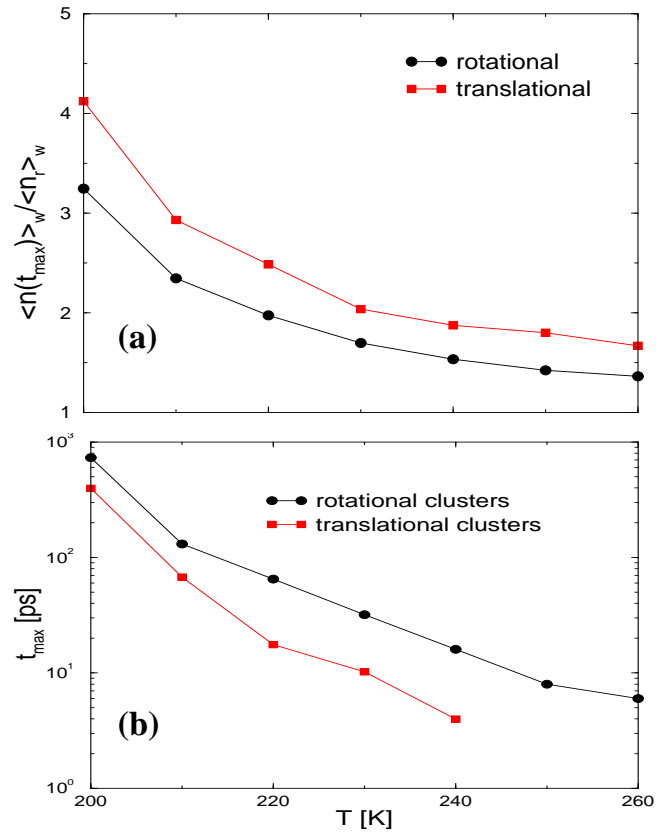}
\end{center}
\caption{(a) Weight average cluster size as a function of temperature
for TH and RH. Both quantities are calculated at the corresponding
$t_{max}$. Note that the TH are larger on average than the RH. (b) The time
$t_{max}$ at which the maximum of the weight average cluster size
occurs. \label{clucomp}}
\end{figure}

Let us now address the question of how these RH depend on the
observation time. Let $\langle n(\Delta t)\rangle$ be the average number
of molecules in a cluster a time $\Delta t$. We define the weight
average cluster size as
\be
\langle n(\Delta t)\rangle_w\equiv\frac{\langle n^2(\Delta t)\rangle}
{\langle n(\Delta t)\rangle}.
\ee
To eliminate the contribution of random clusters, we normalize $\langle
n(\Delta t)\rangle_w$ by $ \langle n_r\rangle_w$, i.e., the weight
average cluster size obtained by chosing randomly $7\%$ of the
molecules.  Figure~\ref{clusize} shows $\langle n(\Delta
t)\rangle_w/\langle n_r\rangle_w$ as a function of $\Delta t$ for
different $T$. In the same manner as the translational case, the maximum
in $\langle n(\Delta t)\rangle_w/\langle n_r\rangle_w$ occurs at the end
of the cage regime of the RMSD, indicating that the cage breaking of the
RDOF is highly correlated with the cage breaking of TDOF. The RH become
larger as $T$ decreases.  Figure~\ref{clucomp}(a) shows $\langle
n(\Delta t)\rangle_w/\langle n_r\rangle_w$ for RH and TH at the
corresponding $t_{max}$, the time at which the corresponding weight
average cluster size is largest. We find that on average TH are larger
than RH, and that RH reach their maximum size before the TH,
Fig.~\ref{clucomp}(b), i.e. $t_{max}$ for RH is smaller than $t_{max}$
for TH.

\begin{figure}
\begin{center}
\includegraphics[scale=0.9]{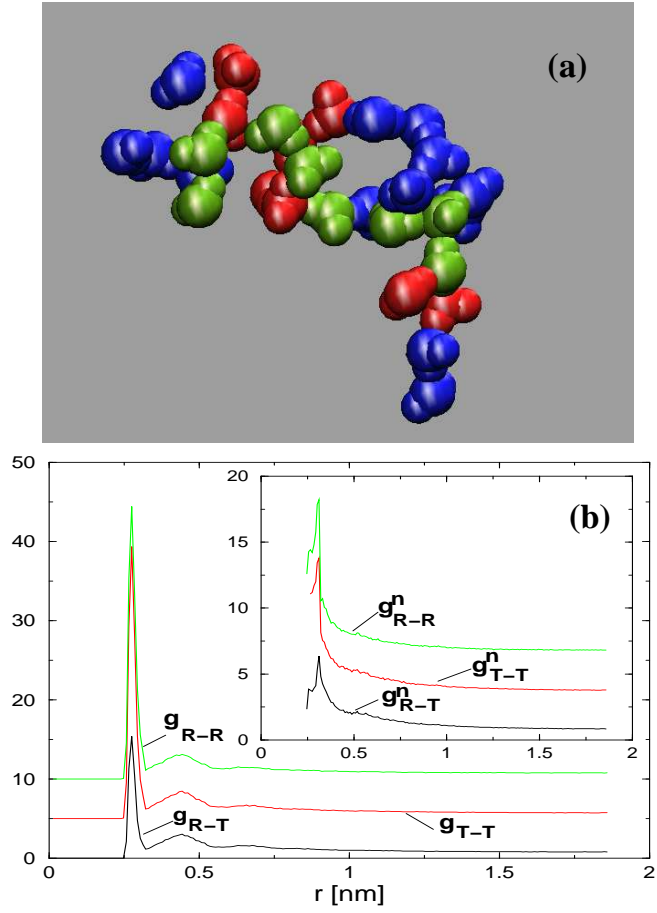}
\end{center}
\caption{(a) A snapshot of one
cluster at $\mathrm{T}=200$ K. The red molecules belong to TH, blue to
RH and the green ones belong to both clusters. (b) RDF between the sets
of translationally mobile and rotationally mobile molecules for
$\mathrm{T}=200$ K. The inset shows the corresponding ratios of these
functions to $g(r)_{bulk}$. For clarity $g_{T-T}$ is shifted by $5$ and
$g_{R-R}$ by 10 on the vertical axis; $g_{T-T}^n$ by $3$ and $g_{R-R}^n$
by $6$ in the inset.\label{crossg}}
\end{figure}

It is natural to ask if RH and TH are formed by the same molecules. We
address this question by simultaneously analyzing the properties of RH
and TH.  Figure~\ref{crossg}(a) is a typical snapshot of the system,
showing both TH and RH. The clusters together form a larger entity
characterizing the dynamical heterogeneities; the molecules belonging to
both kinds of clusters act as the ``backbone'' of such an entity. We
find that the fraction of molecules simultaneously belonging to both
clusters increases with decreasing temperature.

To compare the structure of the TH and RH, we evaluate the radial
distribution function (RDF) of oxygen atoms within each cluster, and
between the two clusters.  Figure~\ref{crossg}(b) shows the RDF for TH,
for RH, and for molecules which belong to \emph{both} TH and RH.  We see that
there is a strong tendency for mobile molecules (of either type) to be
neighbors.  The RDF's are qualitatively similar to the bulk RDF,
$g_{bulk}(r)$, with maxima at $r \approx 0.28$~nm and $r\approx 0.45$~nm
(i.e., molecules are nearest or next-nearest neighbors); however, the
amplitudes of the first peak are strongly enhanced compared to bulk
water.  In order to examine deviations from the bulk we normalize the
RDF's by $g_{bulk}(r)$ [inset of Fig.~\ref{crossg}(b)].  All of the RDF's
display maxima at $0.32$~nm, indicating that oxygens in TH and RH have
an enhanced tendency (with respect to the bulk) to be in the first
interstitial shells of each other and, therefore, have more than four
nearest neighbors. Molecules with five or more neighbors have bifurcated
bonds and represent ``defects'' in the tetrahedral network
characterizing water \cite{bibond,texeira}.  Therefore, our results
suggest that the combined TH and RH in water (i) are a consequence of
the defects of the HB network, and (ii) are primarily composed by
molecules located at the defects of the HB network: evidence of the
relation between the structure and dynamics in water.  Thus the physical
picture needed to describe rotational heterogeneities resembles that
needed for describing translational heterogeneities.

\eject

\begin{thebibliography}{100}
\bibitem{note1} Note that for $t\to 0$, contrary to \cite{nico} for
TDOF, $\alpha_2(t\to 0)\approx 0.37$. This value of $\alpha_2$ can be
obtained analytically from a Boltzmann distribution of a free rotator
and noting that for short times
$\vec{\varphi}_i(t)\simeq\vec{\omega}_i(t)\,\delta t$, where
$\vec{\omega}_i$ is the angular velocity for the $i-$th molecule.
\bibitem{note2} Our results are not qualitatively effected by the choice
of $f$, provided $f \lesssim 15\%$.
\end{thebibliography}
\end{document}